\documentclass[aps,prl,twocolumn,superscriptaddress,showpacs,amsmath]{revtex4}

\usepackage{graphicx,amssymb}

\begin{document}

%

\title{Field-induced magnetoelastic instabilities in antiferromagnetic molecular wheels}

\author{O. Waldmann}
\email[Corresponding author.\\E-mail: ]{waldmann@iac.unibe.ch}
\affiliation{Department of Chemistry and Biochemistry, University of
Bern, 3012 Bern, Switzerland}

\author{C. Dobe}
\affiliation{Department of Chemistry and Biochemistry, University of Bern, 3012 Bern, Switzerland}

\author{S. T. Ochsenbein}
\affiliation{Department of Chemistry and Biochemistry, University of Bern, 3012 Bern, Switzerland}

\author{H. U. G\"udel}
\affiliation{Department of Chemistry and Biochemistry, University of Bern, 3012 Bern, Switzerland}

\author{I. Sheikin}
\affiliation{Grenoble High Magnetic Field Laboratory, CNRS, 38042 Grenoble, France}

\date{\today}

\begin{abstract}
The magnetic torque of the antiferromagnetic molecular wheel CsFe$_8$ was studied down to 50~mK and in fields
up to 28~T. Below ca. 0.5~K phase transitions were observed at the field-induced level-crossings (LCs).
Intermolecular magnetic interactions are very weak excluding an explanation in terms of field-induced
magnetic ordering. A magneto-elastic coupling was considered. A generic model shows that the wheel structure
is unconditionally unstable at the LCs, and the predicted torque curves explain the essential features of
the data well.
\end{abstract}


\maketitle

%

\begin{figure}[b]
\includegraphics[scale=0.8]{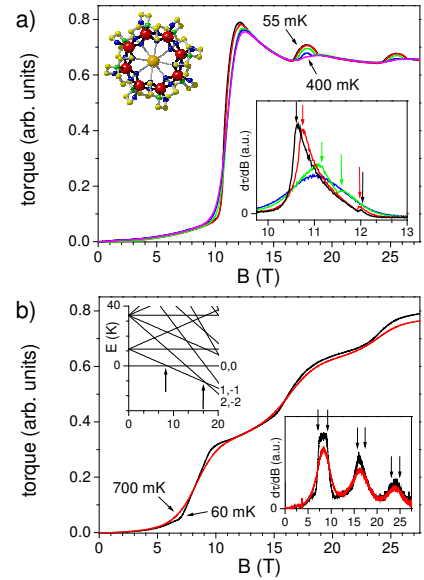}
\caption{\label{fig1} a) Torque vs. field, $\tau(B)$, for a single crystal of {\bf 2} at 55, 100, 200, 300,
and 400~mK (online colors: black, red, blue, green, magenta) ($\varphi$ = -3.3$^\circ$). The right inset
shows $d\tau/dB$ at the first LC, arrows indicate the onset of anomalies. The left inset shows the structure
of CsFe$_8$. b) $\tau(B)$ for a single crystal of {\bf 1} at 60 and 700~mK (online colors: black, red)
($\varphi$ = 93.6$^\circ$). The right inset shows the data as $d\tau/dB$. The left inset shows a calculated
energy spectrum as function of field neglecting magnetic anisotropy ($J = -20.6$~K, arrows mark the first two
LCs, the involved levels are labeled by $S,M$).}
\end{figure}

Antiferromagnetic (AF) molecular wheels attracted huge attention recently because of their peculiar quantum
properties \cite{Taf94,Gat94,Chi98,Cr8,NVT}. These molecules are characterized by a ring-like arrangement of
magnetic metal ions; the ferric wheel [CsFe$_8$\{N(CH$_2$CH$_2$O)$_3$\}$_8$]Cl, or CsFe$_8$ \cite{Saa97},
studied in this work is shown in Fig.~\ref{fig1}(a). The ions within a wheel experience AF nearest-neighbor
Heisenberg interactions, and the molecule's ground state at zero magnetic field is nonmagnetic with total
spin $S = 0$. The next higher lying states belong to $S= 1, 2,$ etc. In a magnetic field these states split
due to the Zeeman interaction, leading to a series of level crossings (LCs) at characteristic fields at which
the ground state changes from the $S=0$, $M=0$ level to the $S=1$, $M=-1$ level, the $S=2$, $M=-2$ level, and
so on, see inset of Fig.~\ref{fig1}(b) \cite{Taf94,Cor99,XFe6}. The magnetization curve at low temperatures
thus exhibits a staircase-like field dependence, with a step at each LC. In this work we report
field-dependent measurements of the magnetic torque on CsFe$_8$, which show clear indications of phase
transitions at the LCs at low temperatures.

The observed anomalies at the LCs could be due to weak magnetic interactions between the molecules in the
sample (\textit{inter}molecular interactions). Since at low temperatures AF wheels behave like dimers
\cite{Cr8}, this would place CsFe$_8$ in the context of weakly-interacting dimer compounds, such as
TlCuCl$_3$, which may exhibit Bose-Einstein condensation of magnons \cite{Nik00,Rue03}. Observation of such
phenomena in a crystal of molecular wheels would be of great interest, but in CsFe$_8$ intermolecular
interactions are very weak excluding such a scenario.

The degeneracy at the LCs suggests a Spin-Peierls type of effect as an alternative, in which the degeneracy
is lifted due to a coupling of the spin system to the lattice. CsFe$_8$ is well described by the spin
Hamiltonian \cite{CsFe8,NVT}
\begin{equation}
 \hat{H} = -J \left( \sum^{7}_{i=1}{ \hat{\textbf{S}}_i \cdot \hat{\textbf{S}}_{i+1} }
                     + \hat{\textbf{S}}_8 \cdot \hat{\textbf{S}}_{1} \right)
  + g \mu_B \hat{\textbf{S}} \cdot \textbf{B} + \hat{H}_A,
\end{equation}
which includes the Heisenberg and Zeeman terms ($J$ = -20.6~K, $g$ = 2), and a term $\hat{H}_A$ describing a
weak uniaxial magnetic anisotropy, mostly due to ligand-field and dipolar interactions (the magnetic
anisotropy is well described by a term $D \sum^8_{i=1} \hat{S}^2_{i,z}$ with $D$ = -0.56~K and the wheel
axis $z$). $\hat{\textbf{S}}_i$ is the spin operator of the $i$th ion with spin $s = 5/2$. Recent
theoretical studies on isotropic AF spin rings ($\hat{H}_A=0$) yielded a magneto-elastic (ME) instability in
zero field for sufficiently small elastic constants, but only for rings with $s = 1/2$ \cite{Spa04,Elh05}.
These conclusions hold also with applied magnetic fields since a modulation of the exchange constants along
the ring cannot lift the degeneracy at the field-induced LCs. However, as will be shown below, the situation
changes markedly in the presence of a magnetic anisotropy, as in CsFe$_8$. A simple, generic model is
introduced, which demonstrates that at the LCs the ring structure is unstable against distortions for any
value of the spin-phonon coupling. The model reproduces the characteristic features of the data well,
suggesting a ME origin of the anomalies at the LCs in CsFe$_8$.

%

Single crystals of CsFe$_8$ were prepared as in Ref.~\onlinecite{Saa97}, but crystallized either from a
mixture of CHCl$_3$ and CH$_2$Cl$_2$ by pentane vapor diffusion yielding
CsFe$_8$$\cdot$5CHCl$_3$$\cdot$0.5H$_2$O (\textbf{1}) \cite{synth}, or from ethanol by diethyl-ether vapor
diffusion yielding CsFe$_8$$\cdot$8C$_2$H$_5$OH (\textbf{2}) \cite{CsFe8}. \textbf{1} (\textbf{2})
crystallizes in the space group P21/n (P4/n) and the molecules exhibit approximate (crystallographic) C$_4$
symmetry. The magnetic parameters, $J$ and $D$, of the molecules are not affected by the solvents
\cite{CsFe8}, but \textit{inter}molecular exchange interactions, if present, should be strongly affected. The
magnetic torque $\tau$ was measured with a CuBe cantilever inserted into the M10 magnet at the Grenoble High
Magnetic Field Laboratory equipped with an Oxford 3He/4He dilution fridge. Background signals were
negligible; only raw data are shown here. In total six single-crystal samples were investigated.

%

Figure~\ref{fig1} presents the field dependence of the torque $\tau$ of CsFe$_8$ at various temperatures for
fields close to the uniaxial axis and almost perpendicular to it ($\varphi$ denotes the angle between field
and uniaxial axis $z$). At the highest temperatures, the curves exhibit the expected behavior: At the LC
fields the torque curves display steps broadened by the effect of, e.g., temperature (a plot $d\tau/dB$ vs.
$B$ shows Gaussian-like peaks with widths $\Gamma$). For CsFe$_8$, the first three LCs are observed in
fields up to 28~T. Normally, the steps become sharper with lower temperatures, corresponding to decreasing
widths $\Gamma$, but in CsFe$_8$, in contrast, a very different behavior is observed at low temperatures.
For fields close to the uniaxial axis, $\varphi \approx 0^\circ$, a dome-shaped contribution to the torque,
centered at the LC fields, appears. This is apparent in Fig.~\ref{fig1}(a) for the LCs at 18~T and 25.5~T,
but also the first LC at 11~T shows the anomaly, as seen in $d\tau/dB$ [inset of Fig.~\ref{fig1}(a)].
Anomalies were also observed for close to perpendicular fields, $\varphi \approx 90^\circ$,
Fig.~\ref{fig1}(b). Here the torque exhibits a linear field dependence at the first LC at 7.7~T, but also at
the higher LCs at 16.5~T and 24~T anomalous behavior is evident from the inset of Fig.~\ref{fig1}(b).

\begin{figure}
\includegraphics[scale=0.8]{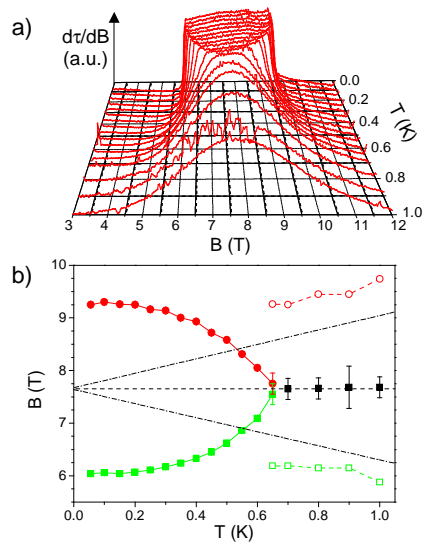}
\caption{\label{fig2} a) $d\tau/dB$ at the first LC for a single crystal of {\bf 1} for several temperatures
from 55~mK to 1~K ($\varphi \approx$ 93$^\circ$). b) $B$-$T$ phase diagram as derived from the data shown in
panel a) (lines are guides to the eyes). The closed symbols indicate the critical fields, the open symbols
the half-maximun fields for $T > T_c$. The dashed line indicates the field of the $S=0 \rightarrow S=1$ LC.
The dash-dotted lines indicate the half-maximum fields as expected for thermal broadening.}
\end{figure}

Figure~\ref{fig2}(a) presents $d\tau/dB$ near the first LC for $\varphi \approx 90^\circ$ as determined from
measurements at several temperatures from 55~mK to 1~K. At the higher temperatures, the curves exhibit the
usual Gaussian-like field dependence, but below a critical temperature $T_c$, of about 0.65~K for the shown
sample, a plateau corresponding to a linear field dependence in $\tau(B)$ emerges. The temperature
dependencies of the lower and upper critical fields, as defined by the kinks in $d\tau/dB$, are plotted in
Fig.~\ref{fig2}(b) (the difference will be denoted as $\Delta B_0$). The figure also displays the maximum and
half-maximum field values of the Gaussian-like curves for $T > T_c$.

All investigated crystals showed the anomalies at the LCs, with a variation in $T_c$ (0.35 to 0.65~K) and
$\Delta B_0$ (1.5 to 3~T). Smaller $\Delta B_0$ corresponded to smaller $T_c$ and vice versa. One origin for
the variation seems to be a dependence of $T_c$ and $\Delta B_0$ on the field orientation indicated by the
data. Within experimental resolution, a hysteresis was not detected.

%

The above torque data clearly demonstrates a phase transition at the LCs in CsFe$_8$. The similarity of the
torque curves in Fig.~\ref{fig1}(b) with magnetization curves observed for systems exhibiting field-induced
magnetic order \cite{Aji89} suggests to assign the anomalies in CsFe$_8$ to intermolecular magnetic
interactions. However, this is an unlikely scenario because i) intermolecular magnetic dipole-dipole
interactions are very weak (even optimistic estimates yield values $<$ 10~mK), ii) in the crystal structures
of \textbf{1} and \textbf{2} the individual molecules are well separated and no exchange pathways exist, iii)
exchange interactions would strongly depend on the solvent and the details of the crystal packing, in
contrast to the observation of similar critical field ranges and temperatures in crystals of \textbf{1} and
\textbf{2}, and iv) no indications of intermolecular interactions were found in the comparable systems
[NaFe$_6$\{N(CH$_2$CH$_2$O$_3$\}$_6$]Cl$\cdot$6CHCl$_3$ down to 0.3~K,
[Fe$_6$\{N(CH$_2$CH$_2$O$_3$\}$_6$]$\cdot$6MeOH down to 0.2~K, and
[NaFe$_6$(OCH$_3$)$_{12}$(C$_{17}$H$_{15}$O$_4$)$_6$]ClO$_4$ down to 40~mK \cite{NaFe6,Pil05,Cin02}.
Intermolecular interactions on the order of several 100~mK are not apparent in CsFe$_8$.

In view of the degeneracy at the LCs a ME effect might be suggested. For isotropic AF spin rings such an
effect is unrealistic, at least for spins $> 1/2$ \cite{Spa04,Elh05}, but, as will be shown below, with an
additional magnetic anisotropy the ring structure becomes unconditionally unstable at the LCs: For a
distorted ring the magnetic anisotropy allows for a mixing of the levels at a LC, resulting in an avoided LC
with a gap $\Delta$ and hence a lowering of the ground-state energy by $\Delta/2$, which is proportional to
the modulation of the anisotropy constants and thus in first approximation \textit{linear} in the distortion.

Following Ref.~\cite{Mn3x3}, the low-$T$ magnetic behavior near a LC is well described by the two-level
Hamiltonian
\begin{equation}
\label{Hss}
 \hat{H}_{S,S+1} = \left( \begin{array}{cc}\epsilon_S & \Delta/2 \\\Delta/2 &
 \epsilon_{S+1}\end{array}\right),
\end{equation}
where $S$ and $S+1$ index the two levels $|S,-S\rangle$ and $|S+1,-S-1\rangle$ involved in the level
crossing. Here, a rotated axis system with the quantization axis parallel to the field $\textbf{B}$ is used,
and $|S,M\rangle$ refers to the labels of the total spin operator in the rotated frame, ${\bf \hat{S}}'^2$
and $\hat{S}_z'$. $\epsilon_S(B,\varphi)$ describes the field and angle dependence of the levels without a
mixing. In first order, one obtains $\epsilon_S(B,\varphi) = - b S + \Delta_S(\varphi)$ with the reduced
field $b = g \mu_B B$, where $\Delta_S(\varphi)$ accounts for the exchange interactions and the zero-field
splittings produced by $\hat{H}_A$ \cite{Cor99,CsFe8lin}. The LC field is given by $b_0(\varphi)=
\Delta_{S+1} - \Delta_S$. A level mixing at the LC is included in the model via $\Delta(\varphi)$, which in
first order is given by $\Delta /2 = \langle S,-S| \hat{H}'_{A} |S+1,-S-1\rangle$ and is thus independent of
the magnetic field ($\hat{H}'_{A}$ is $\hat{H}_A$ expressed in the rotated reference frame).

The symmetry of a non-distorted ring (C$_8$ symmetry of the spin Hamiltonian) prohibits a mixing of the
levels, hence $\Delta = 0$ \cite{Mei01,FW_QT}. A structural distortion induces a modulation of the exchange
and anisotropy constants, which affects both $\Delta_S$ and $\Delta$ [the ME coupling due to $\Delta_S$
($\Delta$) will be called diagonal (non-diagonal)]. $\Delta_S$ is not affected in first order by these
modulations as it is only sensitive to the average of the exchange and anisotropy constants \cite{CsFe8}.
Thus, $\Delta_S$ varies as $\Delta_S \propto x^2$, where $x$ is a parameter describing the structural
distortion (as usual it has been assumed that the modulation of the parameters is linear in the distortion).
This is in accordance with Refs.~\cite{Spa04,Elh05}, and leads to similar conclusions. For the non-diagonal
ME coupling, however, since it is sensitive to the amplitude of the modulations, one finds $\Delta \propto x$
[we write $\Delta(x) = \alpha x$ with the ME coupling constant $\alpha$]. The gain in magnetic energy is now
linear in the distortion resulting in an unconditional ME instability \cite{AJTE}. In this model, the
microscopic details of the non-diagonal ME coupling are lumped into the parameter $\alpha$. Questions
concerning, e.g., the relevant distortion mode thus remain unanswered.

\begin{figure}
\includegraphics[scale=0.8]{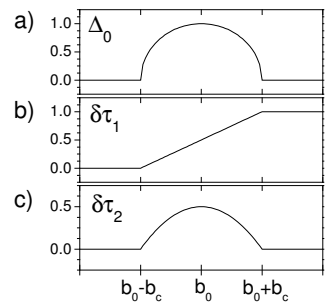}
\caption{\label{fig3} Field dependence of a) the order parameter $\Delta_0$, b) the torque contribution
$\delta \tau_1$, and c) the torque contribution $\delta \tau_2$ around a LC at $b_0$ [the plots were
normalized to $\alpha^2/(2k)$, $\partial b_0 / \partial\varphi$, and $-\partial \Delta_0 / \partial\varphi$,
respectively].}
\end{figure}

The change in ground-state energy due to an opening of a gap at the LC is easily calculated. Including the
elastic energy, the potential $V(x)$ of the total system is
\begin{equation}
\label{Vx}
 V(x) = -{1 \over 2} \sqrt{ (b-b_0)^2 + \Delta(x)^2 } + {1 \over 2}|b-b_0| + {1 \over 2} k x^2,
\end{equation}
where $k$ is the spring constant. In the adiabatic approximation, which neglects the kinetic energy of the
phonons (the validity of this approach is discussed in Ref.~\cite{Elh05}), the equilibrium distortion $x_0$
is given by the minimum of $V(x)$, yielding the condition
\begin{equation}
\label{D}
 (b-b_0)^2 + \Delta_0^2 = \left({\alpha^2 \over 2 k}\right)^2
\end{equation}
with $\Delta_0 = \Delta(x_0)$. Accordingly, as a function of magnetic field, the order parameter $\Delta_0$
describes a semi circle of radius $\alpha^2/(2k)$ around $b_0$, see Fig.~\ref{fig3}(a), and the system
exhibits a spontaneous distortion for fields in between $b_0 \pm b_c$, with $b_c = \alpha^2/(2k)$.

The model permits an analytical calculation of the torque profiles at low temperature. Inserting
eq.~(\ref{D}) into eq.~(4) of Ref.~\cite{Mn3x3}, the change of the torque $\delta \tau$ for magnetic fields
in the critical region $b_0 \pm b_c$ is obtained as
\begin{equation}
\label{tau}
 \delta \tau(b,\varphi) = {1\over 2} {\partial b_0 \over \partial\varphi}
 \left[1 + {(b-b_0) \over b_c}\right] - {1 \over 2} {\partial \Delta_0 \over \partial\varphi} {\Delta_0 \over
 b_c}.
\end{equation}
The torque signal consists of two contributions. The first term, $\delta \tau_1$, describes a linear increase
of the torque from zero at $b \leq b_0-b_c$ to $\partial b_0 / \partial\varphi$ at $b \geq b_0+b_c$
[Fig.~\ref{fig3}(b)], while the second term, $\delta \tau_2$, is proportional to the square of the order
parameter, $\delta \tau_2 \propto \Delta_0^2$ [Fig.~\ref{fig3}(c)]. The relevance with the experimental data
is obvious: the dome-shaped torque contribution produced by $\delta \tau_2$ resembles the behavior at the LCs
for $\varphi \approx 0^\circ$, Fig.~\ref{fig1}(a), and the slope-like contribution of $\delta \tau_1$ that
for $\varphi \approx 90^\circ$, Fig.~\ref{fig1}(b).

In general the torque signal in the critical region is a combination of a slope-like and a dome-shaped curve,
with angle dependent relative weights. The dependence of the LC field on $\varphi$ turns out to be
$b_0(\varphi) \propto 3\cos^2\varphi-1$ \cite{Cor99,CsFe8}, so that the contribution of the slope-like part
to the torque, which is controlled by the factor $\partial b_0 / \partial\varphi$, varies as $\cos\varphi
\sin\varphi$. The variation of the contribution of the dome-shaped part with angle is more complex, since the
various magnetic anisotropy terms of possible relevance may result in very different angle dependencies of
$\Delta_0$. An anisotropy term $\hat{H}_D = \sum D_i [\hat{S}_{iz}^2 - S_i(S_i+1)/3]$, for instance, gives
rise to $\Delta \propto \sin(\varphi) \cos(\varphi)$. A Dzyaloshinsky-Moriya interaction $\hat{H}_{DM} = \sum
\mathbf{d}_i (\hat{\mathbf{S}}_i \times \hat{\mathbf{S}}_{i+1})$, on the other hand, which is likely to arise
in the course of a structural distortion because of the lowered symmetry of the ring \cite{Aff02}, varies as
$\sin(\varphi)$, so that $\partial \Delta_0 /\partial \varphi$ is important at small angles but negligible
near 90$^\circ$. This could explain a dominance of the dome-shaped contribution for parallel fields, and of
the slope-like contribution for perpendicular fields, as observed. The details are not yet understood, and
more studies are clearly needed. However, the proposed scenario is capable of explaining the different
findings for nearly parallel and perpendicular fields, which in our opinion supports the idea of a ME origin
of the observed anomalies in CsFe$_8$.

With $k$ = 10~N/m (corresponding to a typical phonon frequency of $\omega$ = 100~cm$^{-1}$ and a reduced mass
$\mu$ = 14), a $b_c$ of about 1~T implies the reasonable value $\alpha$ = 3~meV/{\AA} (in spin-Peierls
systems, e.g., the coupling constant is $\approx 10 |J| / d_0$ with $d_0 \approx$ 3.5~{\AA}). Concerning the
stability of the distortions against thermal fluctuations, the critical temperature may be estimated from
$1/2 (\partial^2 V/\partial x^2) x_0^2 \approx 1/2 k_B T_c$, i.e., $k_B T_c \approx b_c/2$. A critical field
of about 1~T then suggests $T_c \approx$ 1~K, which is on the order of the observed values. Here however it
should be noted that for systems, which exhibit unconditional lattice instabilities, BCS type of relations
between order parameter and $T_c$ are found \cite{Elh05}. Also, strain effects between the molecules in the
crystal might result in cooperativity which would help to stabilize the distorted phase \cite{CJTE}. A
realistic theory thus might have to include not only the optic but also the acoustic phonons.

%

In conclusion, we have studied the field dependence of the magnetic torque for the AF molecular wheel
CsFe$_8$ and observed anomalies at the level-crossing fields at low temperatures. With respect to their
explanation, several mechanisms were considered. Magnetic interactions between different molecules are very
weak and are thus unlikely to cause the anomalies. As a second possibility, magneto-elastic instabilities
were discussed. Indeed, by introducing a generic model, it has been shown that a non-diagonal
magneto-elastic coupling due to a magnetic anisotropy induces structural instabilities at the LCs. The
predicted torque curves allowed to explain the generic features of the experimental data, in particular the
unusual dome-shaped parts for magnetic fields close to the uniaxial axis. The current work necessarily could
not answer all questions, and more experimental as well as theoretical work is needed. Of most importance
would be structural measurements. On the one hand, they would allow a direct test of the above scenario, and
on the other hand, would yield information about the relevant distortion modes - a crucial input for any
microscopic model to be developed.

For a number of ferric wheels, evidence for gaps at the LCs has been reported \cite{Cor99,Aff02,NaFe6,Cin02}.
In these works, thermodynamic data were found to be better explained by assuming avoided LCs (with
temperature-independent gaps). In this context it is interesting to note that the anomalies observed in
CsFe$_8$ below $T_c$ are announced by an excessive broadening of the torque steps above $T_c$, see
Fig.~\ref{fig2}(b). This mimics avoided LCs, and it will be thus interesting to see whether the earlier
reports of avoided LCs were not in fact first hints of the above anomalies.

%

\begin{acknowledgments}
Financial support by EC-RTN-QUEMOLNA, contract n$^\circ$ MRTN-CT-2003-504880, and the Swiss National Science
Foundation is acknowledged.
\end{acknowledgments}

%

%

\begin{references}
\bibitem{synth}
Details of synthesis and structure will be given elsewhere.
\bibitem{AJTE}
A formal analogy of the model with the Jahn-Teller effect should be noted, see e.g. Ref.~\onlinecite{CJTE}.
This suggests to call it a spin-Jahn-Teller effect.
\bibitem{CsFe8lin}
For CsFe$_8$, because of its rather large $D/J$ ratio, the linear approximation for the field dependence
does not provide truly quantitative results (a quadratic field term would have to be included), but the
model qualitatively grasps all essential features correctly.
\end{references}
\end{document}